\documentclass[12pt, draftclsnofoot, onecolumn]{IEEEtran}
\IEEEoverridecommandlockouts
\usepackage{cite}
\usepackage{amsmath,amssymb,amsfonts,bbm}

\usepackage{textcomp}
\usepackage{xcolor}
\usepackage{comment}
\def\BibTeX{{\rm B\kern-.05em{\sc i\kern-.025em b}\kern-.08em
    T\kern-.1667em\lower.7ex\hbox{E}\kern-.125emX}}
\newcommand\mydots{\hbox to 0.75em{.\hss.\hss.}}

\usepackage{ textcomp }
\usepackage{subfigure}
\usepackage{bm}
\usepackage{multirow}
\usepackage{longtable}
\usepackage{amssymb}
\usepackage{graphicx}
\usepackage{xcolor}
\usepackage{algorithm,algorithmic}
\usepackage[algo2e,ruled,vlined]{algorithm2e}

\newcommand{\edc}[1]{\textcolor{purple}{#1}}

\begin{document}

\title{Distributed Receivers for Extra-Large Scale MIMO Arrays: A Message Passing Approach}
\author{Abolfazl Amiri, Sajad Rezaie, Carles Navarro Manch\'on, Elisabeth de Carvalho 
\thanks{A. Amiri, S. Rezaie, C.N. Manch\'on and E. de Carvalho  are with the Department of Electronic Systems, Aalborg University, 9220 Aalborg, Denmark (e-mail:
{aba,sre,cnm,edc}@es.aau.dk).
}
}
\maketitle

\begin{abstract}
We study  the design of receivers in extra-large scale MIMO (XL-MIMO) systems, i.e. systems in which the base station is equipped with an antenna array of extremely large dimensions. While XL-MIMO can significantly increase the system's spectral efficiency, they present two important challenges. 
One is the increased computational cost of the multi-antenna processing. 
The second is the variations of user energy distribution over the antenna elements and therefore
spatial non-stationarities in these energy distributions.
Such non-stationarities limit the performance of the system.
In this paper, we propose a distributed receiver for such an  XL-MIMO system that can address both challenges. Based on variational  message  passing  (VMP),
We propose a set of receiver options providing a range of complexity-performance characteristics to adapt to different requirements.
Furthermore, we distribute the processing into
local processing units (LPU), that can perform most of the complex processing in parallel, before sharing their outcome with a central processing unit (CPU). Our designs are specifically tailored to exploit the spatial non-stationarities and require lower computations than linear receivers such as zero-forcing.
Our simulation study, performed with a channel model accounting for the special characteristics of XL-MIMO channels, confirms the superior performance of our proposals compared to the state of the art methods. 
\end{abstract}


\section{Introduction}
Massive multiple-input multiple-output (MIMO) systems are known to have high spectral and energy efficiencies that make them a candidate for beyond fifth-generation (B5G) and 6G technologies \cite{Bjornson2019}. Scaling up the number of antenna elements helps getting better performance, as it allows for spatially multiplexing a large number of users on the same time-frequency resources.
Recently, the concept of extra-large scale  MIMO  (XL-MIMO) systems \cite{xlmimo_GC}, or large intelligent surfaces (LIS) \cite{hu2018beyond}, has drawn attention among the researchers. Such systems can provide very high spatial resolutions leading to better quality of services for the mobile users.

One of the main obstacles limiting the possibility of increasing the dimensions of the MIMO array is the computational complexity cost. Most of the well-known conventional linear processing methods such as   zero-forcing  (ZF)  and  minimum  mean  squared  error  (MMSE)  receivers  have prohibitive complexity due to large matrix inversions  when a large number of users is jointly served. Hence, there  is  a  need  for  developing smarter  multi-user  detection  algorithms that deal better with the higher number of  users and antennas at the BS.

According to the electromagnetic propagation effects, in an array with very large dimensions  spatial non-wide sense stationary properties appear \cite{xlmimo_mag}. 
In particular, the large dimension of the antenna array results in a mean energy received from a given user that may vary significantly across the array elements. On the one hand, the large separation between some of the array elements implies that their distance to the user of interest may be significantly different \cite{xlmimo_GC}. On the other hand, array elements that are distant from each other may experience a notably different propagation environment towards the user of interest \cite{gao2013massive}. These effects have been accounted for in proposed channel models by considering visibility regions (VRs) of a given user in the array. The VR for a given user is the subset of BS array elements that hold most of the user's received energy \cite{xlmimo_mag}.
 Presence of VRs limits the system performance compared to the conventional massive MIMO systems where VR sizes are bigger or equal to the array size \cite{anum_nons}. Results in \cite{li2015capacity} also confirm system capacity reduction due to the existence of partially visible clusters in the massive MIMO channel. On the other hand, this property can be useful to design smarter receivers that only consider the processing of the signals received inside the VRs \cite{xlmimo_GC} \cite{amiri2020deep}.

In the following, we investigate the previous works dealing with the XL-MIMO systems and low-complexity receivers. Then, we summarize the contributions of this paper.


    \subsection{ Literature review }

We designed a receiver for large scale MIMO systems in \cite{xlmimo_GC} using distributed units, called \textit{sub-arrays}. First, the central processing unit (CPU) uses a method to assign users to the sub-arrays via a bipartite graph. Then, the sub-arrays detect users' signals by cooperation. We also used a successive interference cancellation (SIC) based method between the sub-arrays, where the effect of each detected user is subtracted from the received signal of the other sub-arrays.
In \cite{rodrigues2020low}, we  used randomized Kaczmarz algorithm (rKA) to design a receiver for XL-MIMO systems where we proposed a heuristic approach to approximate the zero-forcing. Authors in \cite{yang2019uplink} tried to propose a sub-array based architecture for XL-MIMO systems maximizing the 
sum achievable spectral efficiency (SE). They also developed different scheduling methods to achieve better performances.

Among the candidates for multi-antenna B5G technologies,  \textit{cell-free massive MIMO} is trying to remove the cell boundaries and have a user-centric approach for the data-transmission \cite{interdonato2019ubiquitous}. While cell-free network covers a wider area with less density of the users, the XL-MIMO systems operate in a dense area with co-located antennas. The problem of distributing the processing tasks remains the same for both \cite{nguyen2017energy}\cite{bjornson2019making}. However, due to the systematical difference of these two technologies, the focus of the optimization problems varies between them. For instance, cell-free systems have considerable computational delay time \cite{vu2019cell}, i.e. the time spent on exchanging information between BSs and the central node, while this delay is negligible in XL-MIMO systems since all the antennas are co-located.

Developing different algorithms to alleviate the computational costs of the massive MIMO systems is one of the hot ongoing topics in the field of wireless communications. Recently, authors in \cite{sanchez2019decentralized} used a \textit{daisy-chain}
architecture and recursive methods for uplink detection and downlink precoding. On the other hand, \cite{sarajlic2019fully} tried to approximate a ZF precoder on a system where antenna
units at the BS are connected in a daisy chain without a central processing unit and only possess local channel knowledge. The performance of these methods highly depends on the stationary conditions of the users' energy distribution over the antenna array. Therefore, processing the signal from each antenna and moving to the other one (order of choosing the antennas) affects all the users in the same way. However, this does not hold in the XL-MIMO systems, because of the spatial non-stationarities 
where an arbitrary order of the antennas will not guarantee the optimality of the proposed solution. Thus,
for a multiuser detector we will need to find the best antenna sequence of each of the users. This indeed makes the algorithm very complex and is in the opposite direction of the main objective of lowering the complexity.
   
Message passing (MP) has been applied to 
 massive MIMO systems for different application. For instance, in \cite{som2011low} authors develop low complexity MP methods for MIMO inter-symbol-interference systems using graphical models. Both channel estimation and data detection problems with one-bit quantized massive MIMO are solved with variational
approximate message passing (VAMP)
 in \cite{zhang2017one}. In order to deal with complexity growth of MP based methods while using higher order modulations, \cite{zeng2018low} suggested an approximate probability updating scheme. This  scheme considers only the most reliable constellation
point during the message passing process.

Recently, an expectation propagation (EP) based solution for symbol detection in XL-MIMO systems was presented in \cite{wang2019expectation}, where a sub-array structure is assumed to model the EP scheme. Finally, in \cite{amiri2019message} we proposed a variational message passing (VMP) technique for multi-user detection in XL-MIMO systems. Motivated by the complexity behaviour of the VMP (which scales almost linearly with the number of antennas and number of users), we managed to have a low-complexity reception  in  crowd  scenarios. Unlike other linear methods such as matched filtering that fail to operate when favorable propagation conditions do not hold, our method performed very well.
 

\subsection{ Contributions}
In this article, we propose a distributed receiver structure based on variational message-passing and sequential interference cancellation for XL-MIMO systems. In our receiver,  local processing units (LPU) are in charge of executing the VMP per each of the sub-arrays. Then, they forward their post-processed signals to the CPU for the final decision. In this method each of the LPUs can operate independently and in parallel. Moreover,  SIC is used to improve the performance of the symbol detection. The CPU fuses information from all the LPUs and, after removing the signal contributions of detected users, it propagates updated signals back to the LPUs. A secondary parameter called \textit{noise precision} is estimated as well. This parameter gives us a good measure of the quality of users' signals at each sub-array and thus helps to schedule the detection order. Our algorithm uses various initialization schemes  guaranteeing the convergence and performance of the receivers.

The contributions of this work can be summarized as follows:
\begin{itemize}
    \item We present a receiver architecture with flexible complexity-performance trade-off, where we offer a set of receivers that can have low, moderate and high complexity. Depending on their level of complexity, the performance ranges from good to close to optimal. Their complexity also scales at a lower rate than the conventional central linear processing methods, such as the ZF, with regard to the number of users. 
    \item Our receiver distributes symbol detection tasks between the CPU and the LPUs, making it possible to parallelize most of the computations.
    \item  We introduce a generalised non-stationary channel model to capture realistic scenarios. We update the double-scattering channel model in \cite{gesbert2002outdoor} by introducing random spatial non-stationarities in it. This model is based on various measurements data and can model several scenarios and is a more accurate representation for XL-MIMO channels than typically used i.i.d models.
    \item We present various VR-aware receivers that account for the spatial non-stationarities of XL-MIMO channels. These receivers work in a more efficient way by obtaining almost the same performance of the centralized methods while keeping the complexity limited. To the best of our knowledge, this is the first work that exploits the non-stationarities to design a receiver obtaining close to optimal performance.
    
\end{itemize}

\subsection{Paper structure}
First, we start with discussing the channel model in an XL-MIMO system considering the random non-stationarities in Section~\ref{sec:system model}. Then, we explain 
the principles of variational inference,
our problem formulation, the distributed VMP scheme and different initialization techniques in Section~\ref{sec : variational_msg_passing}. After designing the message passing structure,  we aim to detect the transmitted symbols in Section~\ref{sec: symbol_detection}. There, we discuss about  different data fusion methods and symbol detection techniques taking place at the CPU. We conclude our paper with the simulations results in Section~\ref{sec: Simulations} and the conclusions section. 
\subsection{Notations}
Capital calligraphic letters $\mathcal{A}$ denote finite sets. The
cardinality of a set is denoted by $|\mathcal{A}|$. $\mathcal{X}\setminus\mathcal{Y}$ is set $\mathcal{X}$ from which set $\mathcal{Y}$ is excluded.
Boldface small $\bm{a}$ and capital $\bm{A}$ letters stand for vector and matrix representations, respectively. $(.)^H$ is matrix conjugate transpose operator. ${\mathbf{I}_s}$  is an identity matrix of size $s\times s$.
$f(x) \propto g(x)$ denotes that $f(x)=ag(x)$ for some positive constant $a$.
 $\mathbb{E}_x$ is the expectation operator over variable $x$ and $\bar{\mathbf{x}}$  is the mean value of $\mathbf{x}$.
 $\mathcal{CN}(x;\bar{x},\sigma^2_x)$ is a circularly symmetric complex Gaussian distribution with variable $x$ and mean and variance of $\bar{x}$ and $\sigma^2_x$, respectively. $\mathcal{U}(a,b)$ is a uniform distribution in $[a,b]$ interval.

\section{System Model}\label{sec:system model}

We consider a narrow-band MIMO system where $K$ single-antenna active users transmit in the uplink to a base station (BS) with $M$ antenna elements. User symbols are denoted with the vector
$\mathbf{x}\in\mathbb{C}^K$ with entries taking values from the complex constellation set $\mathcal{A}=\{a_1,a_2,\cdots ,a_{|\mathcal{A}|}\}$. $\mathbf{H}=[\bm{h}_1,\cdots, \bm{h}_K]\in\mathbb{C}^{M\times K}$ is the channel matrix and has  column vectors $\bm{h}_k$ each of which denote the channel for user $k$ . At the BS, the noise is assumed to have circularly symmetric complex Gaussian distribution $\mathbf{n}\sim \mathcal{CN}(0,\sigma_n^2\mathbf{I}_M)\in\mathbb{C}^{M}$ ($\mathbf{I}_M$ denotes the identity matrix of size $M$). We model the received baseband signal $\mathbf{y}\in\mathbb{C}^{M}$ across the whole array as follows:
\begin{align}\label{eq:general_model}
\mathbf{y}=\mathbf{H}\mathbf{x}+\mathbf{n}.
\end{align}
The BS is made of a set of $B$ sub-arrays each with $M_b=\frac{M}{B}$ antennas. Here, we define $\Tilde{\mathbf{H}}_b\in\mathbb{C}^{M_b\times K}$ and $\mathbf{y}_b\in\mathbb{C}^{M_b}$ as the channel matrix and the received signal in the $b$-th sub-array for $b\in\{1,\cdots,B\}$, respectively.

\subsection{Channel Model}

\begin{figure*}
	\centering
	\includegraphics[width=0.9\linewidth,trim={0.4cm 0.2cm 0.2cm 0.6cm },clip]{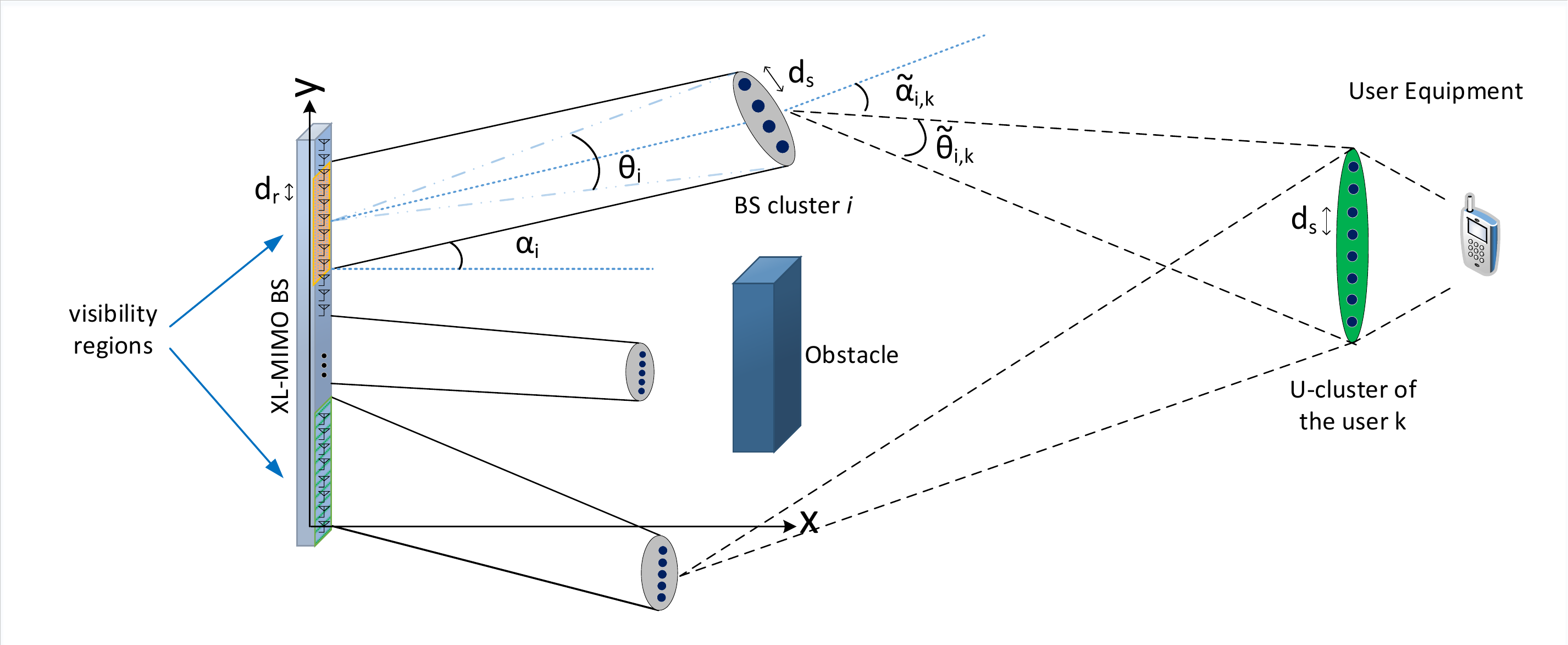}
	\caption{ \small An illustrative example of the propagation model in a XL-MIMO array. Spatial non-stationarities are appearing along the array where most of users' energies are concentrated in the VRs.
	There are several local cluster at the BS side and one cluster per user at each user's side. The interaction between these clusters determines characteristics of the non-stationarities. 
	}
	\label{fig:ex1}
\end{figure*}  

Fig.~\ref{fig:ex1} shows the channel model for the XL-MIMO system accounting for the non-stationary properties of the propagation environment. We adopt a specific channel model called \textit{double-scattering} model \cite{gesbert2002outdoor}. In this channel, correlation is allowed at both the transmitter and the receiver side. In the double scattering channel model, there are two types of scattering clusters in the propagation environment: the one located at the BS side called  \textit{BS-cluster} and one located at the user side called  \textit{U-cluster}. Signals emitted by the users first impinge on
the U-cluster, which scatters them towards multiple BS-clusters directing the signal to the BS array. 

As mentioned before, unlike conventional massive MIMO arrays, XL-MIMO arrays can have a large number of antennas spanning hundreds of wavelengths in space. By decomposing the propagation channel into scatterers, it is observed that the scatterers are not visible over the whole array. This causes variations in the received energy on the array and therefore the appearance of the spatial non-stationarities \cite{xlmimo_mag}\cite{gao2013massive}. 
The main difference between this model
and the model in \cite{gesbert2002outdoor} is that we have several BS-clusters due to the large array size at the BS.
Each of these clusters is seeing a subset of the antennas. 
Furthermore in our model, unlike the previous models proposed in the XL-MIMO literature, the impact
of the spatial fading correlation is decoupled. This indeed will allow us to model  wider ranges of dynamics in our MIMO channel.

The channel between the user $k$ and the BS is modeled as
\begin{align}\label{eq: user k channel}
    \mathbf{h}_k=\left[\tilde{\bm{h}}_{1,k},\cdots,
    \tilde{\bm{h}}_{C,k}\right]\bm{D}_k\:\mathbf{g}_k
    \in\mathbb{C}^{M\times 1}
    ,\;\forall\; k\in\{1,\cdots,K\}
\end{align}
where $\tilde{\bm{h}}_{i,k}\in\mathbb{C}^{M\times S_i}$ denotes the sub-channel for the $i$th BS-cluster with $S_i$ scatterers, $\bm{D}_k\in\{0,1\}^{S'\times S}$ is in charge of assigning the visible BS-clusters to the U-cluster, and the entries of
$\mathbf{g}_k\sim \mathcal{CN}(0,1) \in\mathbb{C}^{S\times 1}$ model the small-scale fading between user $k$ and the $S$ scatterers in  its U-cluster. Moreover, we assume that there are $C$ BS-clusters in the propagation channel and $S'$  scatterers at the BS side visible to the U-cluster.
We formulate the sub-channel of the $i$th BS-cluster and the U-cluster associated to user $k$  as
\begin{align}\label{eq: BS-U-channel}
    \tilde{\bm{h}}_{i,k}=\bm{\Upsilon}_i\bm{\rho}_i^{\frac{1}{2}}\bm{R}_i^{\frac{1}{2}}\bm{G}_i\tilde{\bm{R}}_{i,k}^{\frac{1}{2}}\quad
    \in\mathbb{C}^{M\times S_i},\;\forall i\in\{1,\cdots,C\}\;\text{and}\; k\in\{1,\cdots,K\}
\end{align}
where $\bm{\Upsilon}_i\in\{0,1\}^{M\times r_i}$
determines indices of the antennas at the BS that are visible to the $i$-th BS-cluster with $r_i$ as the number of visible antennas.
$\bm{\rho}_i\in\mathbb{C}^{r_i\times r_i}$ is the visibility gain matrix, $\bm{R}_i\in\mathbb{C}^{r_i\times r_i}$ and 
$\bm{G}_i\in\mathbb{C}^{r_i\times S_i}$
are the correlation matrix and the complex scattering amplitudes between the BS and the BS-cluster $i$, respectively. Also,
$\tilde{\bm{R}}_{i,k}\in\mathbb{C}^{S_i\times S_i}$ is the correlation matrix between the $i$-th BS-cluster and the U-cluster for user $k$. In the following we will discuss each of the channel components in \eqref{eq: user k channel} and \eqref{eq: BS-U-channel} in detail.

\subsubsection{Cluster VR and power distribution}

We have two types of VRs in our propagation channel: cluster VR and user VR. The cluster VR is defined as an antenna region on the BS array that is visible to a cluster. On the other hand, the user VR is a set of the clusters that are being seen by a user. The antenna region for the cluster VR has a random center $c_i$ ( indicating position of the center of the VR in meters) and a random length $l_i$ consisting of $r_i$ consecutive antennas on the BS in the interval
 $[c_i-\frac{l_i}{2},c_i+\frac{l_i}{2}]$ with $d_r$ as the BS antenna spacing. These antennas are belonging to a index set of $\mathcal{R}_i=\{a_1^i,\cdots,a_{r_i}^i\}$ where $a_j^i$ is the $j$-th antenna element inside the $i$-th cluster VR starting from $a_1^i=\lfloor\frac{c_i-\frac{l_i}{2}}{d_r}\rfloor $ ($|\mathcal{R}_i|=r_i$). Note that for two-dimensional arrays, the cluster VR will be an area on the array. We have $C$  cluster VRs that
can  overlap.
Concerning the size of the array, the BS-clusters are all partially visible. This indicates that none of them can see all of the antennas at the BS or in other words, $|\mathcal{R}_i|\neq M$ $\forall \:i\in\{1,\cdots,C\}$. We indicate the relation between the BS antennas and each of the BS-clusters with the antenna association matrix $\bm{\Upsilon}_i\in\{0,1\}^{M\times r_i}$ obtained by:
\begin{align}\label{eq: antenna association matrix}
    \bm{\Upsilon}_i=\left[
    \begin{matrix}\vspace{-0.2em}
    \bm{0}_{m^i_1\times r_i}\\
    \bm{I}_{r_i}\\
     \bm{0}_{m^i_2\times r_i}
    \end{matrix}\right]\qquad,\; m^i_1=a^i_1-1 \;\text{and}\; m^i_2=M-a^i_{r_i}
\end{align}
where $\bm{0}_{m^i\times r_i}$s are zeros matrices and the identity matrix $\bm{I_{r_i}}$ starts
from the $a_1^i$-th row, i.e. from the first antenna of the VR. For example, in a system with $M=5$ antennas and a VR cluster covering antennas in $\mathcal{R}_1=\{2,3,4\}$, this association matrix is
\begin{align}\nonumber
 \bm{\Upsilon}_1 =\left[
    \begin{matrix}\vspace{-0.2em}
    0&0&0\vspace{-0.5em}\\\vspace{-0.5em}
    1&0&0\\\vspace{-0.5em}
    0&1&0\\\vspace{-0.5em}
    0&0&1\\\vspace{-0.2em}
    0&0&0
    \end{matrix}\right]  
\end{align}
The reason for using this matrix is to map the antenna elements into the individual sub-channels for each of the BS-clusters.

The energy distribution is not constant inside each of the cluster VRs. In order to model the variations in the received energy from cluster $i$ within its VR, we refer to the measurements in \cite{gao2013massive}. According to these measurements, energy peak happens at $c_i$ and then it attenuates linearly, i.e with a constant slope $\psi_i$ (dB/m), in a logarithmic scale per distance unit inside the VRs.
We call the energy distribution \textit{visibility gain} and can be calculated using a discrete function such as:
\begin{align}\label{eq: VR gain}
    \rho_i[n]=
              \begin{cases} 10^{-\psi_i|c_i-d_r(n-1)|}\qquad & n\in\mathcal{R}_i\\
              0 & n\notin\mathcal{R}_i
              \end{cases}
\end{align}
where $n$ is the index of the antenna elements inside the cluster VR. We assume a BS array  with antennas located on the y-axis starting from the origin ( see Fig.~\ref{fig:ex1}).
A small portion (less than $5\%$) of the reflected energy from the clusters is spread outside of the VR. For the sake of simplicity in our model, we assume that this energy is zero.
 Measurements in  \cite{gao2013massive} suggest a uniform distribution for $c_i$ along the BS array, a normal distribution for the slope of the gains $\psi_i$ and a log-normal distribution for the cluster VR size $l_i$.
 The measured parameters for these distributions that are used for the simulations are presented in \ref{sec: simulation parameters}. 
Knowing the energy variations inside the VRs, we define the visibility gain matrix $\bm{\rho}_i=\text{diag} (\rho_i[a^i_j]|a^i_j\in\mathcal{R}_i)\in\mathbb{C}^{r_i\times r_i}$ that stores the visibility gains for all of the antennas inside the $i$-th VR in its diagonal entries.



\subsubsection{BS-Clusters}
Aiming to model the correlation between the elements on the receiver side ( at the BS antennas), we define 
the correlation matrix for the BS-cluster $i$ $\mathbf{R}_i\in\mathbb{C}^{r_i\times r_i}$.
This matrix can be calculated as the following where its $(m,l)$ element is \cite{gesbert2002outdoor}:
\begin{align}\label{eq: R_BS_cluster}
   [\mathbf{R}_i]_{m,l}=
   \frac{1}{S_i}\sum_{n=\frac{1-S_i}{2}}^{\frac{S_i-1}{2}}
   \exp\left(-2\pi j (m-l)d_r\cos(\pi/2+\alpha_{i}+\frac{n\theta_{i}}{S_i-1})\right)\quad \forall\;m,l \in \{1,\cdots r_i\}
\end{align}
where, $\theta_{i}$ is the
angular spread and $\alpha_{i}$ is the azimuth angle between the BS and the $i$-th BS-cluster. We assume that the BS-clusters are located randomly in the $x-y$ plane and
the x-axis is the referral line for all the azimuth angles in this paper. 
Moreover, the small-scale fading modelling the complex scattering amplitudes
between the BS and the $i$-th BS-cluster is $\mathbf{G}_i\in\mathbb{C}^{r_i\times S_i}$ and each of its i.i.d entries follow a complex Gaussian distribution $\mathcal{CN}(0,1)$.

In order to indicate the indices of the scatterers in all the BS-clusters, we use an index set.
Without loss of generality, we assume that the scatterers of all the BS-clusters are indexed and stored in scattering set $\mathcal{S}'=\{1,2,\cdots,S'\}$ with $S'=\sum_{i=1}^C S_i$.

\subsubsection{U-Cluster $k$}

This cluster models the scatterers around the user equipment (UE) such as buildings, cars and trees.
We assume $S$ scatterers at the U-cluster. These scatterers are viewed as an array of $S$ virtual antennas that have an average spacing of $d_s$. 
The correlation matrix between the $i$-th BS-cluster and the U-cluster for user $k$ is $\tilde{\mathbf{R}}_{i,k}\in\mathbb{C}^{S_i\times S_i}$.
The $(m,l)$ element of this matrix
is computed as:
\begin{align}\label{eq: R_U_cluster}
     [\tilde{\mathbf{R}}_{i,k}]_{m,l}=
   \frac{1}{S_i}\sum_{n=\frac{1-S_i}{2}}^{\frac{S_i-1}{2}}
   \exp\left(-2\pi j (m-l)d_s\cos(\pi/2-\tilde{\alpha}_{i,k}+\frac{n\tilde{\theta}_{i,k}}{S_i-1})\right)\quad \forall\;m,l \in \{1,\cdots S_i\}  
\end{align}
where, $\tilde{\alpha}_{i,k}$ is the azimuth angle between the $i$-th BS-cluster and the U-cluster of user $k$ and $\tilde{\theta}_{i,k}$ is the corresponding angular spread between them.

\subsubsection{The visibility matrix}

Due to the randomness and obstacles in the environment, 
only a subset of the $C$ BS-clusters ( See Fig.~\ref{fig:ex1}) are visible to the U-cluster $k$.
We denote the user VR by $\mathcal{V}_k$ for each user $k$ containing the indices of the BS-clusters visible for the U-cluster $k$.
We denote the scatterer visibility set for user $k$ as $\mathcal{S}_k\subset \mathcal{S}'$ that stores all the indices of the scatterers of the BS-clusters in $\mathcal{V}_k$.
Now, We can define the visibility matrix  $\bm{D}_k\in\{0,1\}^{S'\times S}$ and its $m$-th row is calculate as:
\begin{align}\label{eq: visibility matrix}
    [\bm{D}_k]_{(m,:)}=
    \begin{cases}
    \bm{1}_{1\times S} \quad &\text{if}\: m\in\mathcal{S}_k\\
    \bm{0}_{1\times S} &\text{otherwise}
    \end{cases}
\end{align}
This matrix shows the visibility of the BS-cluster scatterers to the U-cluster.
As an example, assume $3$ BS-clusters each with $3$ scatterers and a U-cluster with $5$ scatterers and $\mathcal{V}_1=\{1,3\}$. Thus, the resulting scatterer visibility set is $\mathcal{S}_1=\{1,2,3,7,8,9\}$ and the visibility matrix is
\begin{align}\nonumber
 \bm{D}_1 =\left[
    \begin{matrix}
    1&1&1&1&1\vspace{-0.5em}\\\vspace{-0.5em}
    1&1&1&1&1\\\vspace{-0.5em}
    1&1&1&1&1\\\vspace{-0.5em}
    0&0&0&0&0\\\vspace{-0.5em}
    0&0&0&0&0\\\vspace{-0.5em}
    0&0&0&0&0\\\vspace{-0.5em}
    1&1&1&1&1\\\vspace{-0.5em}
    1&1&1&1&1\\\vspace{-0.2em}
    1&1&1&1&1
    \end{matrix}\right]  
\end{align}

Finally, we can assemble the channel for each user $k$ in \eqref{eq: user k channel} using different channel components from equations \eqref{eq: BS-U-channel}-\eqref{eq: visibility matrix}. The scatterer visibility set of user $k$, $\mathcal{S}_k$,  is different and independent from the rest of the users. Thus, users can share some of the BS-clusters depending on these sets. Eventually, the complete channel matrix can be calculated as  $\mathbf{H}=[\bm{h}_1,\cdots,\bm{h}_K]$.

\section{Variational Message Passing}\label{sec : variational_msg_passing}
In this section, we introduce the basics of the VMP method before formulating the problem of estimating the transmitted symbols. Finally, we derive the messages for the VMP algorithm.

\subsection{Variational Inference and Variational Message Passing}

Let $p(\boldsymbol{z}, \boldsymbol{x})$ denote a joint probability density function (pdf), where $\boldsymbol{z}=\{z_1, \dots, z_N\}$ denotes a set of unobserved variables and $\boldsymbol{x}$ a set of observed variables.
Our goal is to find estimates of the variables in $\boldsymbol{z}$ from their marginal posterior pdfs $p(z_i|\boldsymbol{x})$. However, finding these pdfs is often too complex or intractable. Instead, we resort to computing a surrogate distribution $q(\boldsymbol{z})$ that approximates the posterior $p(\boldsymbol{z}| \boldsymbol{x})$ and from which marginals $q_i(z_i)$ can be easily found.
This is obtained in variational inference by minimizing their Kullback-Leibler (KL) divergence, defined as
\begin{align}
    D(q||p)\triangleq \int q(\boldsymbol{z}) \log \frac{q(\boldsymbol{z})}{p(\boldsymbol{z}|\boldsymbol{x})}dz.
\end{align}
To make the problem tractable, the surrogate function $q$ is restricted to fulfill certain constraints. Typically, the mean-field approximation, which considers a fully factorized distribution of the form
\begin{align}
    q(\boldsymbol{z})=\prod_{i=1}^N q_i(z_i)
\end{align}
is applied, in addition to normalization constraints $\int q_i(z_i)dz_i=1$. With these constraints, a sequential minimization of the KL divergence with respect to each of the factors $q_i$ is performed. It can be shown \cite{Bishop} that, at each step, the optimal factor $q_i$ given all other factors $q_j, j\neq i$ is obtained by
\begin{align}\label{eq:MFupdate}
    q(z_i) \propto \exp\left(\mathbb{E}_{j\neq i}\{\ln{p(\boldsymbol{z},\boldsymbol{x})}\}\right)
\end{align}
where the expectation is taken with respect to all approximate marginals $q_j, j\neq i$. The above update rule is applied alternately to the different factors until convergence is achieved.

This algorithm can also be formulated in terms of a message passing algorithm. Assume that the joint distribution factorizes as
\begin{align}\label{eq:joint_pdf}
    p(\boldsymbol{z},\boldsymbol{x}) = \prod_a f_a(\boldsymbol{z}_a,\boldsymbol{x}_a)
\end{align}
where $\boldsymbol{z}_a\subseteq \boldsymbol{z}$ and $\boldsymbol{x}_a\subseteq \boldsymbol{x}$ are subsets of the unobserved and observed variables.
All of the $f_a(\boldsymbol{z}_a,\boldsymbol{x}_a)$s are the factors in the joint pdf and they depend on the statistical dependencies in the model. The factorization is not unique, as several factors can be combined together. Moreover, this factorization can be graphically represented as a factor graph which we will introduce it in the next subsection.
The update in \eqref{eq:MFupdate} can be expressed in terms of messages passed along the edges of the factor graph \cite{Bishop} as in 
\begin{align}
    q_i(z_i) \propto \prod_{f_a\in \mathcal{N}(z_i)} m_{f_a \to z_i} (z_i)
\end{align}
where $\mathcal{N}(z_i)$ denotes the set of factors in \eqref{eq:joint_pdf} that contain variable $z_i$, and the messages read
\begin{align}\label{eq:VMP_msg_format}
    m_{f_a \to z_i} (z_i) = \exp\left(\mathbb{E}_{j\neq i}\{ \ln f_a(\boldsymbol{z}_a,\boldsymbol{x}_a) \}\right).
\end{align}

\subsection{Probabilistic System Description}

To apply the VMP inference described above, we first formulate a probabilistic model of the system. Our ultimate goal is to infer the values of the transmitted symbols $x_k$, $k=1,\dots,K$ and of the unknown noise precision (inverse of the noise variance) $\lambda=\frac{1}{\sigma_n^2}$. Ideally, this should be done from their joint posterior distribution, which reads
\begin{align}
    p(\mathbf{x}, \lambda|\mathbf{y}) \propto p(\mathbf{y}|\mathbf{x}, \lambda)p(\mathbf{x})p(\lambda)
\end{align}
where, due to the white Gaussian noise, $p(\mathbf{y}|\mathbf{x}, \lambda)=\mathcal{CN}(\mathbf{y}; \mathbf{Hx}, \frac{1}{\lambda}\mathbf{I}_M)$. The prior symbol distribution reads $p(\mathbf{x})=\prod_k p(x_k)$, with $p(x_k)$ being uniform over the constellation set $\mathcal{A}$, and we assume the noise precision to have a non-informative Gamma prior. 

\begin{figure}
	\centering
	\includegraphics[width=0.48\linewidth]{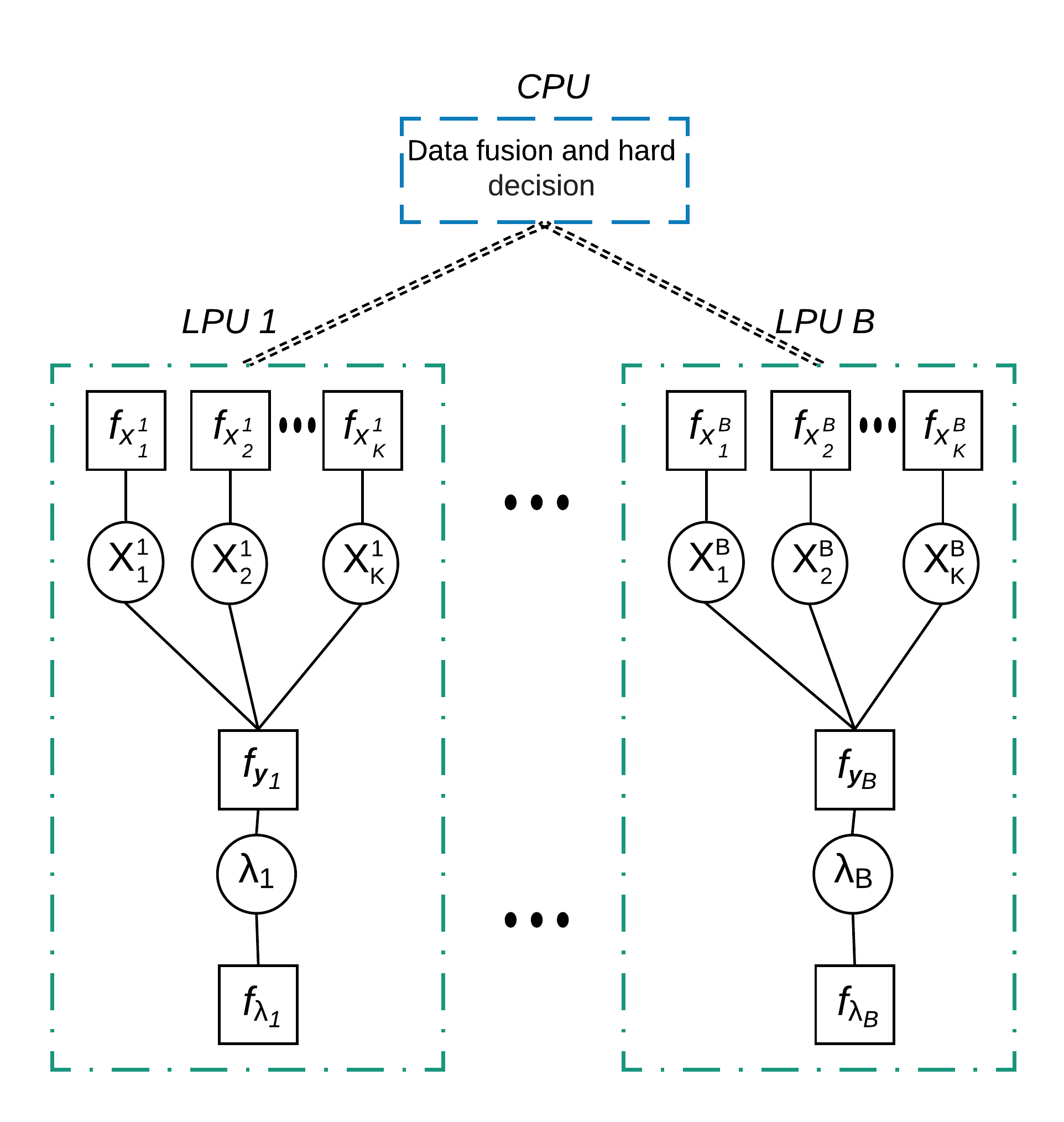}
	\caption{ \small Factor graph representation of the local processing units and the central unit. Local estimations obtained by the local units are sent to the central unit for the data fusion and detection. 
	}
	\label{fig:ex2}
\end{figure}

As we are interested in performing as much of the processing locally at each of the BS sub-arrays, however, we formulate instead $B$ similar models, one for each of the sub-arrays:
\begin{align}\label{eq:joint_prob}
    p(\mathbf{x}^b,\lambda_b|\mathbf{y}_b)\propto \underbrace{p(\mathbf{y}_b|x_1^b,\cdots,x_K^b,\lambda_b)}_{f_{\mathbf{y}_b}}\underbrace{p(\lambda_b)}_{f_{\lambda_b}}\prod_{k=1}^K\underbrace{p(x_k^b)}_{f_{x_k^b}}, \qquad b=1,\dots,B.
\end{align}
where the variables $\mathbf{x}^b=[x_1^b,\dots,x_K^b]^T$ and $\lambda_b$ denote respectively the transmitted symbols and noise precision observed by the $b$-th BS sub-array, $f_{\mathbf{y}_b}(\mathbf{x}^b, \lambda_b)=\mathcal{CN}(\mathbf{y}_b; {\tilde{\mathbf{H}}_b\mathbf{x}^b}, \frac{1}{\lambda_b}\mathbf{I}_{M_b})$, $f_{x_k^b}(x_k^b)=\frac{1}{|\mathcal{A}|}\bm{1}(x_k^b\in\mathcal{A})$, and $f_{\lambda_b}\propto 1/\lambda_b$ \footnote{This choice of prior corresponds to an improper, noninformative Gamma prior distribution with shape and rate parameters approaching zero.}. Although, clearly, $\mathbf{x}^b$ and $\lambda_b$ represent the same random variables for the different sub-arrays $b=1,\dots,B$, we treat them separately here such that each sub-array can, in their local processing phase, obtain independent estimates of them based on solely their received signals $\mathbf{y}_b$. After each local processing phase, their respective estimates are fused in a CPU and distributed back to the sub-arrays. The factor graphs illustrating the models in \eqref{eq:joint_prob} and their linking with the CPU are depicted in Fig.~\ref{fig:ex2}.

\subsection{VMP at the Local Processing Units}

We proceed in this section to describing the VMP algorithm run at each of the LPUs at the sub-arrays. The $b$-th LPU aims at approximating the posterior in \eqref{eq:joint_prob} by using the approximate distribution
\begin{align}
    q_b(\mathbf{x}^b,\lambda_b) = q_{\lambda_b}(\lambda_b)\prod_{k=1}^Kq_{x_k^b}(x_k^b)
\end{align}
where the na\"\i ve mean-field approximation is applied. At each of the local processing rounds, an initial setting for the factors $q_{x_k^b}(x_k^b)$, $k=1,\dots,K$ is available, with different initialization strategies discussed in Section~\ref{sec:initilization}. In the first step, the message from factor node $f_{\mathbf{y}_b}$ to the variable node $\lambda_b$ is calculated as
\begin{align}\label{eq:VMP_basic}
   m_{ f_{\mathbf{y}_b\longrightarrow {\lambda_b}}}({\lambda_b})\propto \exp{(\mathbb{E}_{\mathbf{x}^b}\{\ln{\big(f_{\mathbf{y}_b}(\mathbf{x}^b,{\lambda_b})\big)}\})}\propto {\lambda_b}^{M_b} \exp (-{\lambda_b} Z_b)
\end{align}
where $\mathbb{E}_{\mathbf{x}^b}\{\cdot\}$ denotes the expectation 
with respect to the initial distribution $q_{\mathbf{x}^b}({\mathbf{x}^b}) = \prod_{k=1}^K q_{x_k^b}(x_k^b)$, and $Z_b={||\mathbf{y}_b-\sum_k \Tilde{\mathbf{h}}_{b,k} \bar{x}_k^b||^2+ \sum_k \sigma^2_{x_k^b}\Tilde{\mathbf{h}}_{b,k}^H \Tilde{\mathbf{h}}_{b,k}}$. In this expression, $\Tilde{\mathbf{h}}_{b,k}$ denotes the $k$th column of $\Tilde{\mathbf{H}}_{b}$ while $\bar{x}_k^b=\sum_{s\in\mathcal{A}}s q_{x_k^b}(s)$ and $\sigma_{x_k^b}^2 = \sum_{s\in\mathcal{A}}|s|^2 q_{x_k^b}(s)-|\bar{x}_k^b|^2$ are the mean and variance of ${x}_k^b$ with respect to $q_{x_k^b}(x_k^b)$.
The approximate marginal distribution $q_{\lambda_b}(\lambda_b)$ is then obtained by multiplying the messages entering the variable node $\lambda_b$ as
\begin{align}\label{eq:q_lambda}
    q_{\lambda_b}\!({\lambda_b})\!\!=\!f_{\lambda_b} ({\lambda_b})\!\times\!  m_{ f_{\mathbf{y}_b\rightarrow {\lambda_b}}}\! ({\lambda_b})=\! {\lambda_b}^{\overbrace{M_b-1}^{\alpha-1}}e^{ -{\lambda_b}\! \overbrace{(\!Z_b)}^\beta}
\end{align}
which correspondes to a Gamma distribution with mean \begin{align}\label{eq:lambda_bar}
      \bar{\lambda}_b =\! \frac{\alpha}{\beta}\!
    =\!\frac{M_b}{Z_b}.
\end{align}

Next, the LPU computes the messages from factor node $f_{\mathbf{y}_b}$ to the variable nodes $x_k^b$, which result in
\begin{align}\label{eq: f_y_to_x}
    m_{f_{\mathbf{y}_b\longrightarrow x^b_k}}\!\!\! \propto \exp{(\mathbb{E}_{\lambda_b, \mathbf{x}_{\setminus k}^b}\{\ln{\big(f_{\mathbf{y}_b}(\mathbf{x}^b,{\lambda_b})\big)}\})} \propto  \mathcal{CN}\bigg(\!x^b_k;\! \frac{\Tilde{\mathbf{h}}_{b,k}^H}{||\Tilde{\mathbf{h}}_{b,k}||^2}
    (\mathbf{y}_b\!-\!\!\!\!\sum_{k'\neq k}\!\bar{x}^b_{k'}\Tilde{\mathbf{h}}_{b,k'}),\! \frac{1}{\bar{\lambda}_b||\Tilde{\mathbf{h}}_{b,k}||^2}\bigg)
\end{align}
where, similarly as in \eqref{eq:VMP_basic}, $\mathbb{E}_{\lambda_b, \mathbf{x}_{\setminus k}^b}\{\cdot\}$ denotes the expectation with respect to $q_{\lambda_b}(\lambda_b)$ and $\prod_{k^{\prime}\neq k}q_{x_{k^\prime}^b}(x_{k^\prime}^b)$.

To finalize, the approximate marginals of the symbols of each user at the sub-array $b$ are obtained by multiplying these messages with their local priors, yielding
\begin{align}\label{eq: SA_qx}
    q_{x_k}^b(x^b_k) \propto m_{f_{\mathbf{y}_b}\longrightarrow x^b_k}(x^b_k)\times f_{x_k^b}(x_k^b).
\end{align}

\subsection{Initialization Options}\label{sec:initilization}

As mentioned above, VMP requires initial approximate symbol distributions $q_{x_k^b}^{0}(x_k^b)$ to begin its operation, which we review next.
\subsubsection{Type of the initialization}
The simplest option is to initialize the algorithm with a uniform distribution where all the symbols are, a priori, equiprobable for all users. In this case, the initial distributions are set as ${q_{x_k^b}^{0}(x_k^b)}=\frac{1}{|\mathcal{A}|}\bm{1}(x_k^b\in\mathcal{A})$,  $\forall k \in \{1,\cdots,K\}$ and $\forall b \in \{1,\cdots,B\}$. This method has no computational complexity, but typically results in slow convergence of the VMP algorithm. 

The performance and convergence speed of the local processing can be improved by using linear processing techniques to set the initial symbol distributions. 
As a first option, we consider maximum ratio combining (MRC) over all the BS array.
Applying the MRC for user $k$ to the received signal in \eqref{eq:general_model} yields
\begin{align}\label{eq:MRC_filter}
    \hat x_k^{\text{MRC}}=\frac{\mathbf{h}_k^H}{||\mathbf{h}_k||^2}\mathbf{y}= x_k + \sum_{k'\neq k}^K\frac{\mathbf{h}_k^H}{||\mathbf{h}_k||^2} \mathbf{h}_{k'} x_{k'}+\frac{\mathbf{h}_k^H}{||\mathbf{h}_k||^2}\mathbf{n}
\end{align}
Assuming a large number of users ($K\gg 1$), the sum of the second and third terms in \eqref{eq:MRC_filter}
can be approximated as a complex Gaussian random variable according to the central limit theory. The initial approximate marginals of the symbols $x_k$'s are therefore set as proportional to a Gaussian pdf, restricted to the symbol alphabet $\mathcal{A}$, i.e.
\begin{align}\label{eq:MRC_init}
    {q_{x_k^b}^{0}(x_k^b)} \!\propto\! \mathcal{CN}\bigg(\!x_k^b; \hat x_k^{\text{MRC}}\!\!\!,\frac{\sum_{k'\neq k}^K P_{x_{k'}}
    |\mathbf{h}_k^H \mathbf{h}_{k'}|^2 +||\mathbf{h}_k||^2\sigma_n^2}
    {||\mathbf{h}_k||^4}\!\bigg)\bm{1}(x_k^b\in\mathcal{A}), \qquad \forall b\in\{1, \dots, B\}.
\end{align}
where $P_{x_k}=\mathbb{E}\{x_kx_k^H\}$ is user signal power. This initialization introduces a complexity load of $3MK$ multiplications.
There is also another possibility to apply the MRC initialization locally and at each of the LPUs. For this type, local channel vectors $\tilde{\bm{h}}_{b,k}$ and $\tilde{\bm{h}}_{b,k'}$ and local received signal $\bm{y}_b$ are used in \eqref{eq:MRC_filter} for each sub-array $b$. Then, local estimates are calculated using \eqref{eq:MRC_init} for all of the $B$ sub-arrays.

Another candidate for the initialization is the ZF method.
The ZF receiver filter for user $k$, denoted by $ \mathbf{F}_{\text{ZF},k} $, reads
\begin{align}\label{eq:ZF_filter}
\mathbf{F}_{\text{ZF},k}[\mathbf{H}]=\frac{\mathbf{h}^H_k\mathbf{P}^{\bot}_{\mathbf{\bar{H}}_k}}{\mathbf{h}^H_k\mathbf{P}^{\bot}_{\mathbf{\bar{H}}_k}\mathbf{h}_k},
\end{align} 
with $\mathbf{P}^{\bot}_{\mathbf{\bar{H}}_k}=\mathbf{I}-\mathbf{\bar{H}}_k(\mathbf{\bar{H}}_k^H\mathbf{\bar{H}}_k)^{-1}\mathbf{\bar{H}}_k^H$ \cite{brown2012practical}; $\mathbf{\bar{H}}_k $ is obtained from $ \mathbf{H} $ by removing its $k^{th}$ column $\mathbf{h}_k$.
The resulting estimates after application of the ZF filter are 
\begin{align}
\hat x_k^{\text{ZF}}&=\mathbf{F}_{\text{ZF},k}\mathbf{y}=x_k+\mathbf{F}_{\text{ZF},k}\mathbf{n}
\end{align}
and they have mean equal to the transmitted symbol $x_k$ and a variance given by
\begin{align}\label{eq:snrzf}
\sigma_{x_k^{\text{ZF}}}^2&=\sigma_n^2\left( \mathbf{h}^H_k\mathbf{P}^{\bot}_{\mathbf{\bar{H}}_k}\mathbf{h}_k\right)^{-1}.
\end{align}
Similarly as for MRC initialization, we approximate the inital marginals as
\begin{align}\label{eq:ZF_init}
    {q_{x_k^b}^{0}(x_k^b)} \!=\! \mathcal{CN}\bigg(x_k^b; \hat x_k^{\text{ZF}},\sigma_{x_k^{\text{ZF}}}^2\bigg)\bm{1}(x_k^b\in\mathcal{A}), \qquad \forall b\in\{1, \dots, B\}.
\end{align}

A last option is to perform similar ZF initialization but applied locally at each of the LPUs. In this case, the ZF filter for the $b$th sub-array is computed analogously to \eqref{eq:ZF_filter} but using channel matrix $\Tilde{\mathbf{H}}_{b}$ instead of $\mathbf{H}$. After this, an approximate marginal similar to that in \eqref{eq:ZF_init} is calculated for each of the $B$ sub-arrays.

\subsubsection{Strategy}
In this subsection, we present two different modes to initialize the VMP method. The first option is to initialize the VMP just a single time.
We call this mode \textit{One-time} initialization.
The second option is to initialize the algorithm multiple times. This mode is done to help stabilizing the outputs within the consecutive iterations of the VMP method. This mode becomes more interesting when we initialize the VMP at each step of the interference cancellation detection. There, after each step of the interference removal, the linear pre-processing used for the initialization will perform more accurately and improve the performance of the overall scheme.

With this, we finalize the description of the processing of the LPUs of each sub-array, which is summarized in Algorithm~\ref{alg1}.

\begin{algorithm}[t]
\small{
	\SetAlgoLined
	\KwResult{Local symbol estimates for all  active users}
	\emph{Initialize:} 
	 $M$,  $K$, $\mathbf{y}$, sub-array index $b$ , parameters in Sec.~\ref{sec:system model}, $\mathcal{A}$, VMP iterations $\mathcal{I}$.
	
1. Get the corresponding channel matrix for sub-array $b$ from $\Tilde{\mathbf{H}}_b$ that is generated using \eqref{eq: user k channel}.

2. Choose one of the initialization methods in (\ref{sec:initilization}) set the initial probabilities as $q^0(x_k)$.

\For{$i = 1$ to $\mathcal{I}$ }{

3. Extract $\bar{x}^b_k$ and $\sigma^2_{x^b_k}$ values from  $q_{x_k}^{(i-1)}(x_k)$. 

4. Calculate the mean value of the precision parameter $\bar{\lambda}_b$ using \eqref{eq:lambda_bar} for the sub-array.

5. Calculate symbol probabilities $q^{(i)}_{x_k}(x_k)$ using \eqref{eq: SA_qx} for all the users $k=\{1,\cdots, K\}$.
	}

6. Finalize the local information to be sent to the CPU $q^{b}_{x_k}(x_k)$=$q^{(i)}_{x_k}(x_k)$.

	\caption{\small VMP at each of the LPUs.}
	\label{alg1}}
\end{algorithm}

\section{Data Fusion and Symbol Detection}\label{sec: symbol_detection}

\begin{figure}
	\centering
	\includegraphics[width=0.7\linewidth]{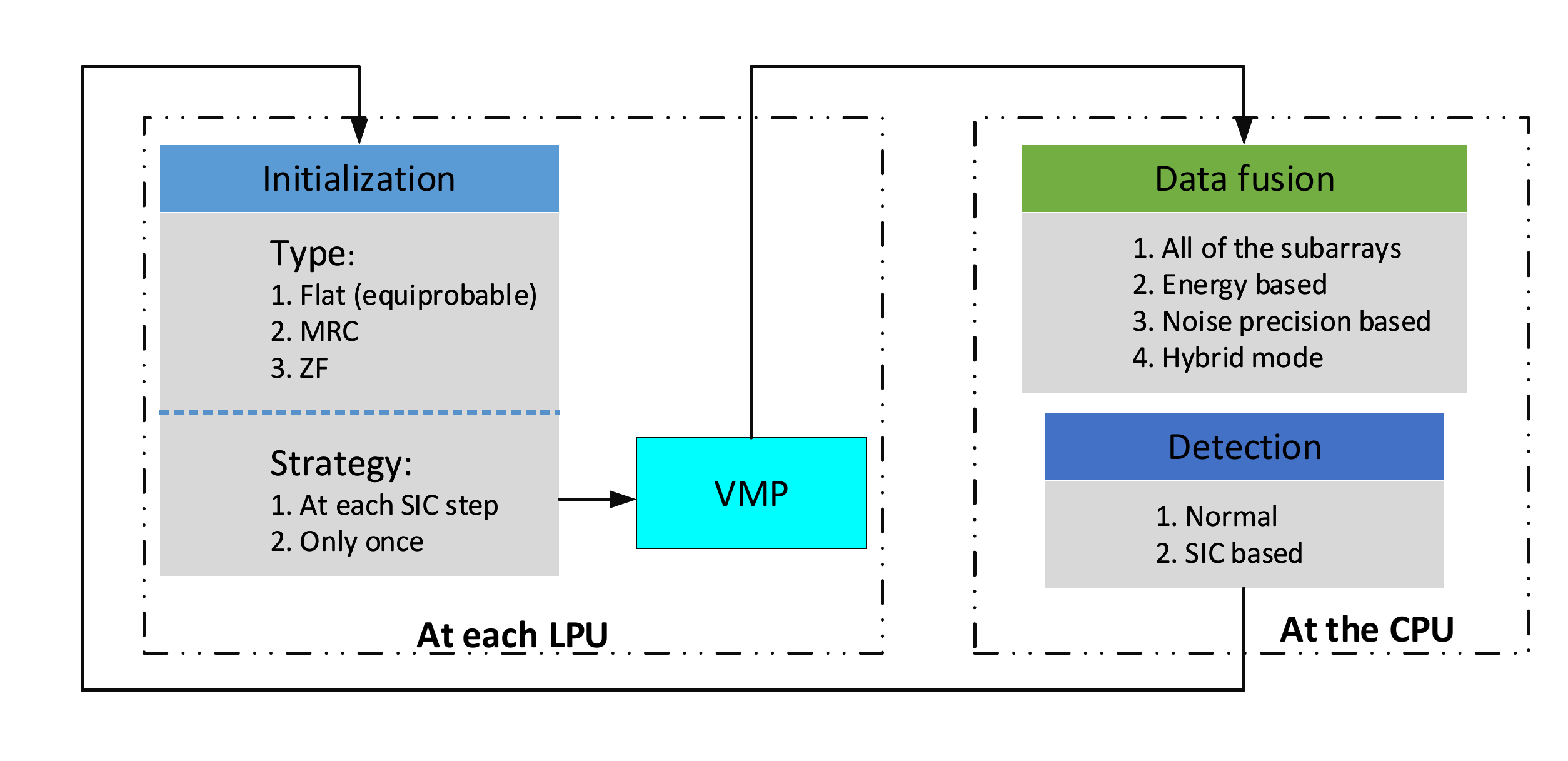}
	\caption{ \small A summary of the algorithms for the VMP based symbol detection in the XL-MIMO system. 
	}
	\label{fig:diag}
\end{figure}

In this section, we detail how the results of the local VMP processing performed by the LPUs at each of the sub-arrays are combined at the CPU to yield the final symbol estimates. 
The overall receiver process is illustrated in the block diagram in Fig.~\ref{fig:diag}. The operations are divided between the LPUs and the CPU while offering each of them several options. At the LPUs illustrated in the left-hand side of the diagram, VMP processing is performed as discussed in Section~\ref{sec : variational_msg_passing}, including the different initialization options. On the right-hand side of the diagram, the CPU is illustrated as having two basic tasks: the fusion of the symbol estimates provided by the different LPUs, and the eventual detection of the symbols by using the fused information. Four different options are studied for the data fusion process, and two options are considered for detection: non-iterative, and SIC based. In the latter case of SIC based detection, several iterations of local and central processing are performed before the symbols of all users are detected, which is illustrated in the diagram by the feedback connection between the CPU and the LPU.
In the following we introduce each of the options. 

\subsection{Data Fusion at the CPU}\label{sec:Data Fusion at the CPU }

After the local processing at each LPU is performed as described in Section~\ref{sec : variational_msg_passing} and Algorithm~\ref{alg1}, the local approximate marginals from all the sub-arrays are sent to the CPU. The CPU fuses the received information to get an overall estimate of each user's symbol.
The data fusion is done based on a sub-array data fusion binary matrix defined as
\begin{align}
    \mathbf{{V}}=\left (\begin{matrix} v_{1,1} & v_{1,2} & \cdots & v_{1,K} \\\vdots & \vdots & \ddots  &  \vdots \\ v_{B,1}  & v_{B,2}& \cdots & v_{B,K}  \end{matrix} \right)\in\{0,1\}^{B\times K}.
\end{align}
If $v_{b,k}=1$ then the local estimate $x_k^b$ of sub-array $b$ will contribute to the global estimate of $x_k$, otherwise if $v_{b,k}=0$ it will not be fused.
The data fusion is done by multiplying the estimates coming from each of the sub-arrays. Thus, the global estimates can be calculated as
\begin{align}\label{eq:data_fusion}
 q_{x_k}(x_k) \propto \prod_{b\in\mathcal{F}_k} q_{x_k^b}(x_k)
\end{align}
where $\mathcal{F}_k = \{b\in\{1,\dots,B\}|v_{b,k}=1\}$, i.e., for the $k$th user symbol the product is only taken over those sub-arrays $b$ whose entry $v_{b,k}=1$.
Depending on the values of $\mathbf{V}$, different data fusion strategies can be selected. The simplest choice of the data fusion is when the CPU fuses data from all of the sub-arrays, i.e. $\mathbf{V}=\bm{1}_{B\times K}$ (a matrix of all ones). 
\begin{algorithm}[H]
\small{
	\SetAlgoLined
	\KwResult{$\mathbf{V}$}
	\emph{Initialize:} $ \mathbf{H},\:K,\:B,\:B_{\text{max}},\:M_b,\: p_0,\:\bm{\lambda},\: \mathbf{V}=\{0\}^{B\times K},\:\mathcal{B}=\left\{1,\hdots,B\right\},data\: fusion\: type$\\
\uIf{$data\: fusion\: type==$PWR}
{
\For{$k = 1$ to $K$ }{
1. reinitialize $p_k=0$

2. compute total cumulative power $P_k=||\mathbf{h}_k||_2^2$

3. compute per sub-array power $P_k^{(b)}=||\Tilde{\mathbf{H}}_{b,k}||_2^2,b\in\mathcal{B}$

	\While{$ p_k\leq p_0\cdot P_k $}{
		1. find $b^*=\max_{b\in\mathcal{B}} P_k^{(b)}$
		
	    2.	$p_k=p_k+P_k^{(b^*)}$
	
	    3.	set $\mathbf{V}(b^*,k)=1$
	
	    4.	$\mathcal{B}=\mathcal{B}\setminus {{b}^{*}}$ 
	
		}}}
\uElseIf{$data\: fusion\: type==$NOP}
{
1. sort $\bm{\lambda}$'s elements decreasingly ($\mathbf{b}=\arg \text{sort}(\bm{\lambda})$)

2. choose the first $B_{\text{max}}$ elements of the sorted indices vector $\mathbf{b}$ as $\mathbf{b}^\ast=\mathbf{b}(1:B_{\text{max}})$

3. set $\mathbf{V}(\mathbf{b}^\ast,:)=1$

} 
\uElseIf{$data\: fusion\: type==$HYB}	
{
1. compute per sub-array power $P_k^{(b)},b\in\mathcal{B}$ and $\mathbf{P}=\{p_{b,k}|p_{b,k}=P_k^{(b)}\: \forall k,b\}$

2. compute the hybrid measure matrix from \eqref{eq:hybrid_matrix} 

\For{$k=1$ to $K$}{
3. sort elements of the $k$-th column of $\mathbf{\Gamma}$ decreasingly ($\mathbf{b_k}=\arg \text{sort}(\mathbf{\Gamma}(:,k))$)

4. choose the first $B_{\text{max}}$ elements of $\mathbf{b_k}$ as $\mathbf{b_k}^\ast=\mathbf{b_k}(1:B_{\text{max}})$

5. set $\mathbf{V}(\mathbf{b_k}^\ast,k)=1$
}
}
\Else{All sub-arrays fusion :
    $\mathbf{V}=\bm{1}_{B\times K}$
  }
	\caption{\small sub-array data fusion matrix $\mathbf{V}$ construction.}
	\label{alg3}}
\end{algorithm}

Other selection options exploiting the non-stationary spatial structure of the received signals over the array are discussed next, with the goal of finding a data fusion matrix $\mathbf{V}$ which yields an advantageous complexity-performance trade-off.

\subsubsection{Power based data fusion}
One of the main differences of an XL-MIMO system with a conventional one is the variations of received users' energy along the array. For instance, measurements in \cite{gao2013massive} confirm that the energy from each user is not evenly distributed along all the elements of the array. Therefore, this property can be exploited to reduce the complexity of the receivers by processing only the parts of the array with the highest energy. 
Furthermore, results in \cite{xlmimo_mag} confirm that processing parts of the XL-MIMO array that contain a significant amount of energy is enough to get almost the same spectral efficiency (SE) as processing all the elements of the array.
Here, we use the same principle by assigning the users to the sub-arrays that contain a certain ratio of the user's energy, for instance $80\%$, as we introduced in \cite{xlmimo_GC}.  
This method is noted by \textit{PWR} mode in Algorithm~\ref{alg3}.


\subsubsection{Noise precision based data fusion}
    As mentioned in Section~\ref{sec : variational_msg_passing}, the noise precision parameter $\lambda_b$ is a measure of residual interference and AWGN variance per each sub-array. Therefore, more reliable sub-arrays for signal detection are those with the highest $\bar{\lambda}_b$ values. In order to limit the complexity, we restrict the algorithm to choose the top $B_{\text{max}}$ sub-arrays.
    This method is included as \textit{NOP} mode in Algorithm~\ref{alg3}, where $\bm{\lambda}=\{\bar{\lambda}_b|\forall b\in \mathcal{B}\}$ and $\mathcal{B}=\{1,\cdots, B\}$ is the set of all sub-arrays.

\subsubsection{Hybrid data fusion}
Inspired by \eqref{eq: f_y_to_x} and the variance of the Gaussian distribution, another possible metric to select a subset of sub-arrays for the data fusion task is to choose the ones with the lowest variance in \eqref{eq: f_y_to_x}. In order to calculate this metric, we define a \textit{hybrid measure matrix} $\mathbf{\Gamma}\in \mathbb{R}^{B\times K}$ as
\begin{align}\label{eq:hybrid_matrix}
   \mathbf{\Gamma}=diag(\bm{\lambda})\mathbf{P} 
\end{align}
where, $\mathbf{P}=\{p_{b,k}|p_{b,k}=P_k^{(b)}\: \forall k,b\}$ is the energy of each user in each of the sub-arrays. Then, based on the number of sub-arrays to be processed $B_{\text{max}}$, we select the top $B_{\text{max}}$ sub-arrays for each of the users. These sub-array subsets determine the data fusion candidates for each of the user symbols. This method is listed by \textit{HYB} mode in Algorithm~\ref{alg3}.

\subsection{Symbol Detection Strategies}

After the data fusion process, CPU can decide how to detect the users' symbols.
In the following, we discuss two different approaches for the detection in the CPU.

\subsubsection{Non-iterative Data Fusion and Detection}

In this case, the CPU fuses all the local estimates and no further processing is done over the fused information.
The CPU demodulates the global estimation for each of the users' transmitted symbol, $q_{x_k}(x_k)$, and detects as  constellation point from the set $\mathcal{A}$ with largest approximate marginal. 
Therefore, the detection is done by the following probability comparison 
\begin{align}\label{eq:symbol_detection}
    \hat{x}_k=\arg\max_{a\in \mathcal{A}} q_{x_k}(a) \quad\forall k
\end{align}
where the symbol maximizing the global estimate is chosen as the detected symbol for each of the users.

\subsubsection{SIC Data-Fusion and Detection}

One of the effective ways to boost the receiver performance is to use SIC. This type of detector works sequentially and at each step detects the strongest user (or layer) and then removes its effect from the received signal. This operation reduces the interference successively and therefore improves the probability of  successful detection of the subsequent symbols. 
One of the main factors that determine the performance of the SIC detector is the user (or layer) ordering method. In our investigation, we applied different ordering criteria based on the post-processing SINR, received users' energy, users' noise precision, etc, but none produced satisfactory results. Instead, We propose a new metric 
called likelihood ratio (LR) metric,
based on the ratio of probabilities  between the top two most likely symbols.
In order to  define the symbol certainty,
first, we sort the symbol probabilities provided by the approximate marginals $q_{x_k}(x_k)$ of each user $k$ as $p^{(k)}_1\geq p^{(k)}_2\geq\cdots p^{(k)}_{|\mathcal{A}|}$.
Next, we define the symbol certainty as
\begin{align}
    {\Delta_k}\triangleq \frac{p^{(k)}_1}{p^{(k)}_2}, \qquad k=1,\dots,K.
\end{align}
Finally, the LR metric 
which chooses the strongest user as
\begin{align}\label{eq:LR_measure}
    k^{\ast}=\arg\max_{k} {\Delta_k}
\end{align}

Algorithm~\ref{alg2} represents the SIC mechanism and cooperation between the CPU and the LPUs. It consists of the following parts at each SIC step:
\begin{itemize}
    \item Algorithm~\ref{alg1} and fusion of the local estimates  $q_{x_k^{b}}(x_k^b)$ using \eqref{eq:data_fusion}.
    \item Selecting the strongest user $k^\ast$ using \eqref{eq:LR_measure}
    \item Detecting $k^\ast$'s symbol as $\Tilde{x}_{k^{\ast}}$ by making a hard decision
    \item Interference cancellation by updating the received signal as $\mathbf{y}\leftarrow \mathbf{y}-\Tilde{x}_{k^{\ast}}\mathbf{h}_{{k^{\ast}}}$
    \item Fixing the prior for $k^{\ast}$ by setting $q_{x_{k^{\ast}}}({x_{k^{\ast}}})=\delta(x_{k^{\ast}}-\Tilde{x}_{k^{\ast}})$\footnote{$\delta(.)$ is the Dirac delta function.}
    \item Sending back the updated prior $q_{x_{k^{\ast}}}$ and $\bm{y}$ to the LPUs
\end{itemize}

\begin{algorithm}[t]
\small{
	\SetAlgoLined
	\KwResult{Central symbol estimates using the SIC detection}
	\emph{Initialize:} 
	 $\mathbf{y}$,  $K$, $\mathcal{A}$, detected symbols set $\mathcal{S}=\phi$.

1. Define user set $\mathcal{K}\triangleq\{1,\cdots,K\}$

\For{$i = 1$ to $K$ }{

2. Run Algorithm~\ref{alg1} to get the local estimates $q_{x_k}^b(x_k)$ for all the LPUs $b\in\{1,\cdots,B\}$.
	
3. Fuse the local estimates to get $q_{x_k}(x_k)$ from \eqref{eq:data_fusion} for all the users in $\mathcal{K}$.

4. Choose the strongest user $k^{\ast}$ to detect with the LR measure using \eqref{eq:LR_measure}.

5. Detect the transmitted symbol for $k^{\ast}$ as $\Tilde{x}_{k^{\ast}}$ and include it to the detected symbol set $\mathcal{S}\leftarrow \mathcal{S} \cup \Tilde{x}_{k^{\ast}} $.

6. Cancel the interference caused by ${k^{\ast}}$ by: $\mathbf{y}\leftarrow \mathbf{y}-\Tilde{x}_{k^{\ast}}\mathbf{h}_{{k^{\ast}}}$ and remove $\mathbf{h}_{{k^{\ast}}}$ from $\mathbf{H}$

7. Fix the prior for $k^{\ast}$ as $q_{x_{k^{\ast}}}({x_{k^{\ast}}})=\delta(x_{k^{\ast}}-\Tilde{x}_{k^{\ast}})$

8. $\mathcal{K}\leftarrow\mathcal{K}\setminus k^{\ast}$

	}

9. \textbf{Output}: $\mathcal{S}$.

	\caption{\small Data fusion and the SIC at the CPU.}
	\label{alg2}}
\end{algorithm}



\section{Performance Evaluation}\label{sec: Simulations}
In this section, we present numerical results for the performance
of the proposed algorithms. We begin by describing the used simulation mode and the selected benchmarks. We follow by analysing the computational complexity of the methods and end by illustrating the performance of the proposed receivers with respect to the benchmarks.

\subsubsection{Generating the channel and simulation parameters}\label{sec: simulation parameters}
Intending to generate the channel model in \eqref{eq: user k channel}, and based the on simulation parameters in Table~\ref{tab:sim_parameters}, first we generate the VRs. After assigning $n_b$ random clusters to each user $k$ and forming $\mathcal{V}_k$, the correlation matrices are calculated using \eqref{eq: R_BS_cluster} and \eqref{eq: R_U_cluster}. Finally, we normalize the generated channel to the value of trace$(\mathbb{E}\{\mathbf{H}\mathbf{H}^H\})$ to ensure unit expected received power per each antenna element at the BS side.\footnote{This normalization corresponds to assuming ideal UE power control.}
Note that, the fast fading parameters are generated at each channel realization while the correlation matrices are updated  every $50$ realizations to model the long term statistics of the channel.

\begin{table}
\centering
\caption{Simulations Parameters}
\label{tab:sim_parameters}
\begin{tabular}{|c|c||c|c||c|c|}
\hline
\bf{Variable} & \bf{Value} & \bf{Variable}& \bf{Value} &\bf{Variable} & \bf{Value}  \\\hline	

$M$    &    $256$  &   $K$   &  $32 $  &
$d_r$   & $0.0578\;m$   \\\hline
$d_S$  & $5\; m$ &
$\mathcal{I}$    &    $1$  &   $|\mathcal{A}|$   &  $4 $ (QPSK)  \\\hline

$C$    &  $20$  & $l_i$ & lognormal$(0.7,0.2)$ &
$c_i$    &  $\mathcal{U}(0,M)$  \\\hline

$\psi_i$ & normal$(-0.21,0.8)$ &
$S_i$   & $31$   & $\alpha_i,\tilde{\alpha}_{i,k}$  & $ \mathcal{U}(-\pi/2,\pi/2)$  \\\hline

$\tilde{\theta}_{i,k}$   & $3\pi/4$   & $\theta_i$  & $7\pi/8$  &
$B$   & $4$ \\\hline $B_{\text{max}}$  & $3$ &

$p_0$  &  $0.75$  &  $n_b$  & $4$

 \\\hline

\end{tabular}

\end{table}

\subsection{Benchmarks}
\subsubsection{Ideal matched filter bound}
 We choose matched filter bound which is the case when the effect of all the other users are ideally canceled and the target user's signal is detected by MRC. This  single-user  detection  in  the  interference-free  channel gives the best achievable SER \cite{burchill1995matched}.
\subsubsection{Central linear processes}
To compare the performance of our proposed receiver with centralized linear processing methods, MRC and ZF benchmarks are implemented. These are obtained by respectively applying \eqref{eq:MRC_filter} and \eqref{eq:ZF_filter}.


\subsubsection{Expectation propagation method from \cite{wang2019expectation}}
Aiming to have a benchmark method for a message passing based scheme, we implemented the EP algorithm presented in \cite{wang2019expectation}. This method works in a distributed manner where  the sub-arrays are exchanging their local estimates and a final decision is taken in the central node. It is worth mentioning that the computational complexity of this method is higher than our method due to the matrix inversions and singular value decompositions required. However, here we only consider the symbol error detection results regardless of the complexity.

\subsection{Complexity Analyses}
In this subsection we analyse the complexity of the aforementioned methods. The complexity for the central  ZF and the central MRC are \cite{bjornson2015optimal}
\begin{align}
    C_{\mbox{\small{ZF}}}&=\frac{K^3}{3}+MK^2+MK\\
    C_{\mbox{\small{MRC}}}&=3MK.
\end{align}
 
In order to calculate the complexity of the VMP method in both of the algorithms, we start with Algorithm~\ref{alg1} and analyse each of the steps separately. Their complexity is reported in  Table~\ref{tab: complexity}.
\begin{table*}
\centering
\caption{Complexity analyses for Alg.~\ref{alg1}}
\label{tab: complexity}
\begin{tabular}{|c||c||c|}
\hline
 \bf{Step} &  \bf{$\#$ of multiplications} & \bf{Remarks}\\\hline	
 1  & None & \textendash\\\hline
 2  & $3M_bK$ & For the MRC initialization   
   \\\hline
 3  & $2K$ & Two operations per each user   
   \\\hline
 4  & $2KM_b+2M_b+M_b$ & Two summations and one $l2$ norm  
   \\\hline
 6  & $K|\mathcal{A}|$ & $|\mathcal{A}|$ constellation points for each user   
   \\\hline
 \textendash  & $\mathcal{I}(K(2+2M_b+|\mathcal{A}|)+2M_b)+3M_bK$ & Total number of multiplications  
   \\\hline
  
\end{tabular}

\end{table*}
Next, the total complexity of Algorithm~\ref{alg2} is calculated using the complexity values obtained for VMP processing at each sub-array. To begin with, we discuss the following remarks regarding this algorithm:
 \begin{itemize}
     \item \textbf{Remark 1}: The number of VMP iterations $\mathcal{I}$ is one of the important parameters in the complexity-convergence performance of the VMP method. We tested different values for $\mathcal{I}$ and found that the VMP converges at $\mathcal{I}=1$ and there is no need to repeat the operations.
    Thus, $\mathcal{I}=1$ is assumed for all simulations and analyses henceforth.
    
     \item \textbf{Remark 2}: At each SIC iteration, the number of undetected users decreases by $1$. Therefore, we have to consider a variable complexity for step $2$ of Alg.~\ref{alg2} due to the size of $\mathcal{K}$. This can be done easily by factorizing $K$ from the expressions in Table~\ref{tab: complexity} as $C_\text{VMP} \approx Kf(M_b,\mathcal{A})$ and a summation over different values of $K$. For instance, algorithm starts with $K$ users, then in the second round with $K-1$ users and so on. The total complexity can be approximated as $\sum_{k=K}^1 kf(M_b,\mathcal{A}) \approx K^2 f(M_b,\mathcal{A})/2 $.
     \item \textbf{Remark 3}: For the VR based VMP methods the complexity depends on $B_{\text{max}}$ and $p_0$ values. Thus, with a rough comparison, the complexity will scale with factor of $B_{\text{max}}/B$ and $p_0$ for noise precision and power based data fusion methods, respectively. For instance, considering $75\%$ power threshold or having $B_{\text{max}}=3$ in a system with $B=4$ will approximately reduce the total complexity by $25\%$.
 \end{itemize}
 
Hence, the total number of multiplications of the Algorithm~\ref{alg2} is 
\begin{align}
    C_{\mbox{\small{SIC-VMP}}}\approx \frac{K^2}{2}(B(5M_b+|\mathcal{A}|+2)+1)+MK
\end{align}
which is a second-order function of $K$. We will compare the numerical evaluations of the expressions we derived in this subsection later in Sec.~\ref{sec: Simulations}.

  \begin{figure}
     \centering
     \includegraphics[width=0.7\linewidth,trim={1.4cm 0 1.6cm 0.6cm },clip]{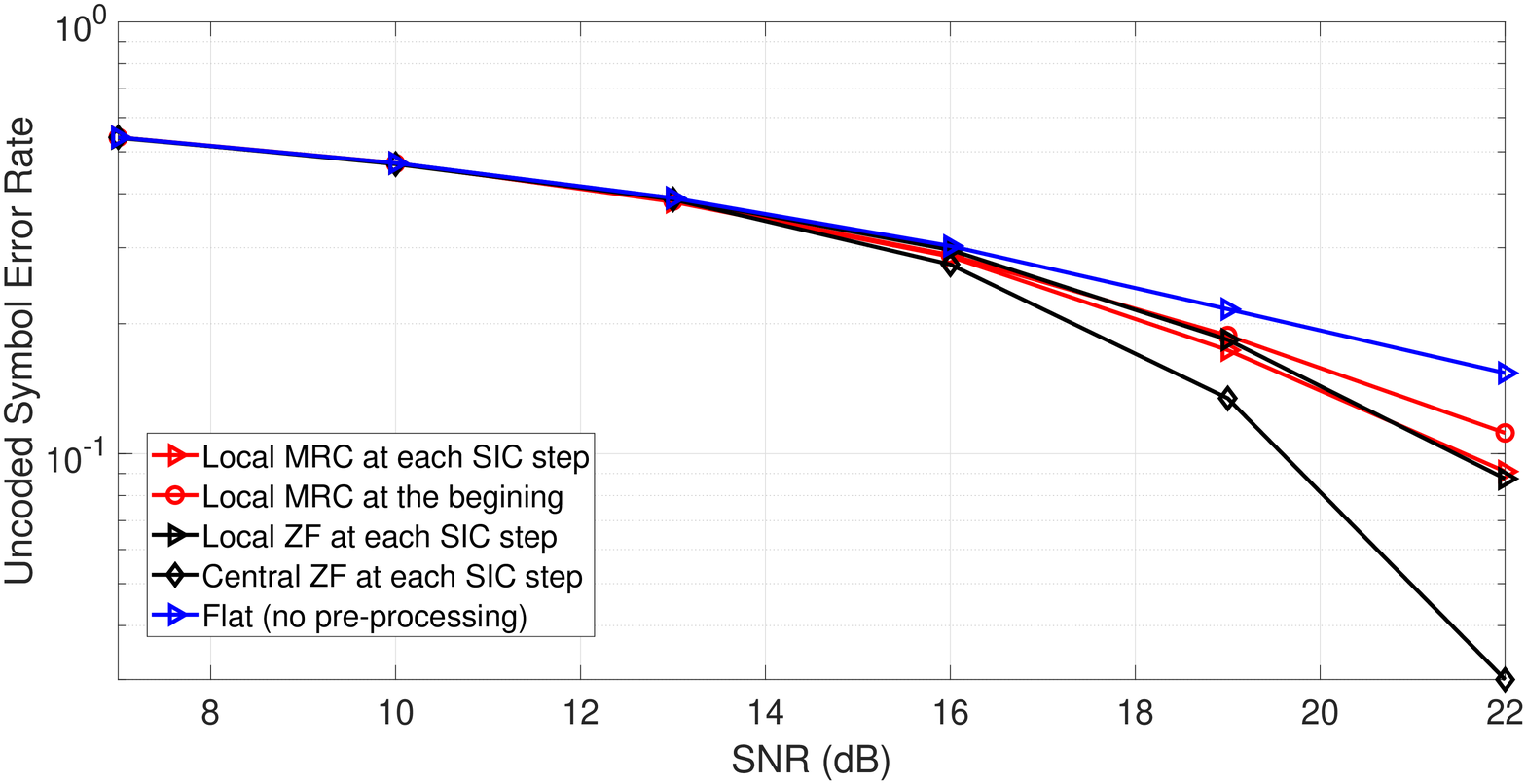}
     \caption{\small Different initialization techniques and their effect on the probability of error. In this simulations we used a lowly-correlated channel having the parameters in Table.~\ref{tab:sim_parameters} with $M=128$, $K=16$, $B=2$.}
     \label{fig:SER_vs_initialization}
 \end{figure}
 
 Fig.~\ref{fig:SER_vs_initialization} shows the different performances depending of the initialization types discussed in \ref{sec:initilization}. As expected,
 the best performance is when we initialize at each step of the SIC operation. Due to the similar performance of the ZF and the MRC methods, it is more favorable to use the MRC mode for its less complexity. Thus, from now on we use the MRC at each step of the SIC as our default initialization method in all of the VMP performance evaluations.

\subsection{Simulation Results}
\label{sec:simulation_results}

\begin{figure*}
\centering
\subfigure[Uncorrelated channel]{
     \centering
     \includegraphics[width=0.5\textwidth,trim={1.4cm 0 1.6cm 0.6cm },clip]{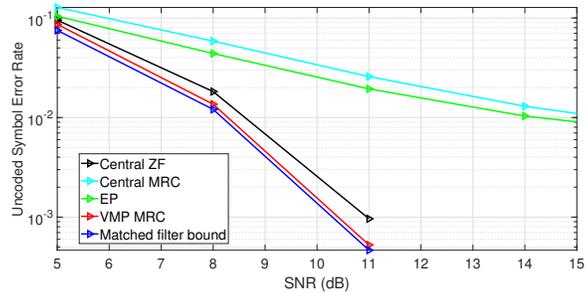}\label{fig:iid_channel}}
     \subfigure[Low-correlated channel]{
     \centering
     \includegraphics[width=0.5\textwidth,trim={1.4cm 0 1.6cm 0.6cm },clip]{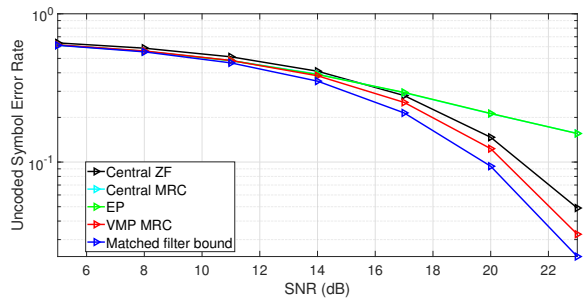}\label{fig:low_channel}
}  
\subfigure[High-correlated channel]{
     \centering
     \includegraphics[width=0.5\textwidth,trim={1.4cm 0 1.6cm 0.6cm },clip]{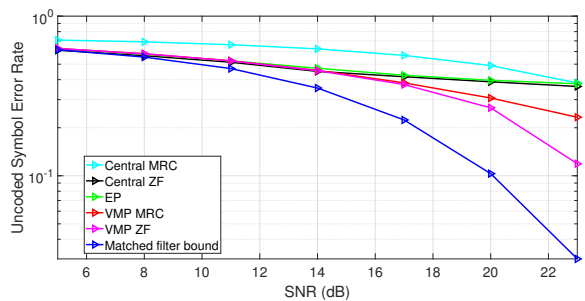}\label{fig:high_channel}
}   
     \caption{\small Uncoded SER of different methods vs pre-processing SNR for different correlation scenarios.}
     \label{fig: sim SER SNR}
 \end{figure*}

     

 \begin{figure}
     \centering
     \includegraphics[width=0.7\linewidth,trim={1.4cm 0 1.6cm 0.6cm },clip]{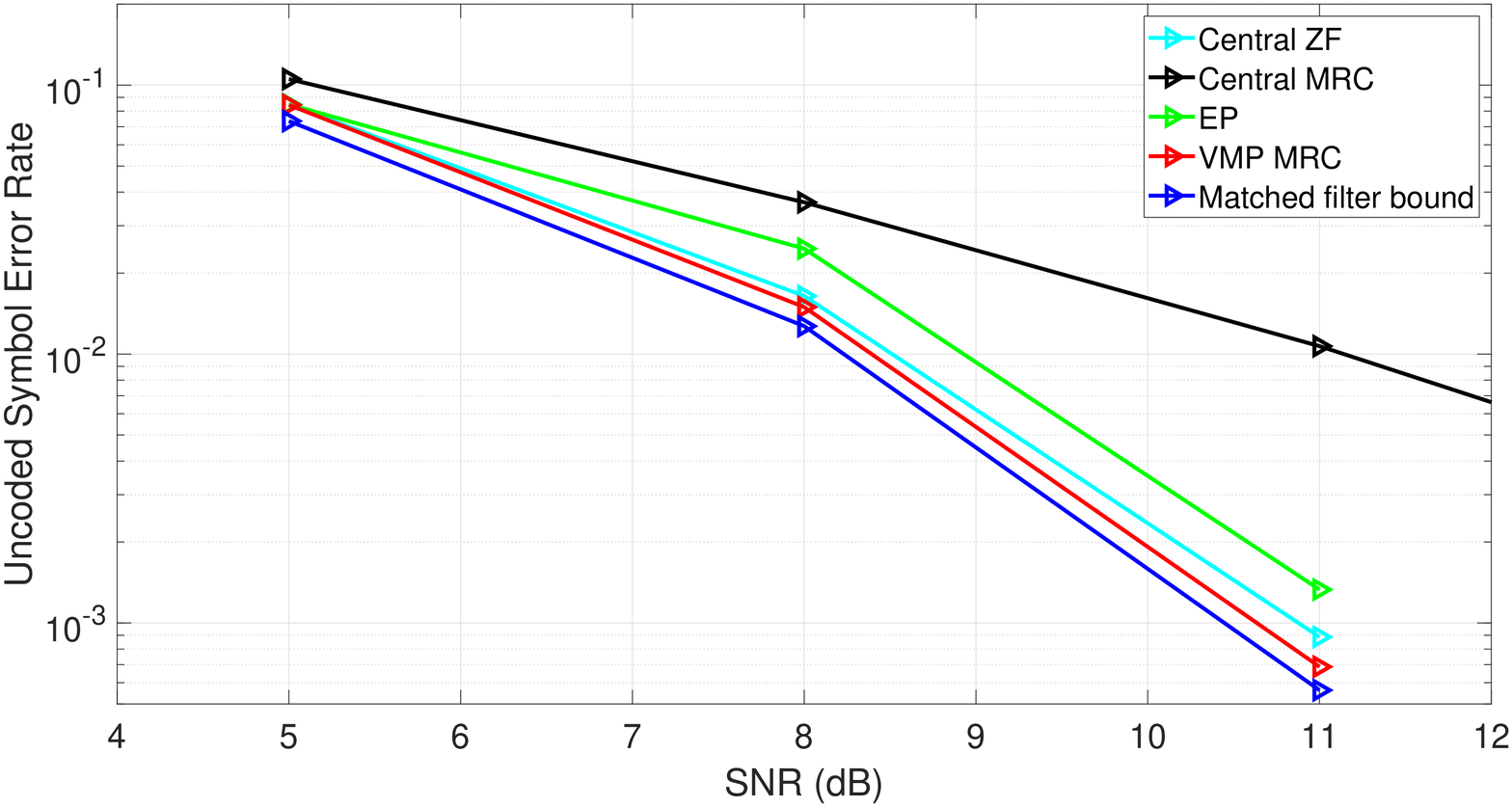}
     \caption{\small Uncoded SER of different methods vs pre-processing SNR. $M=100$, $K=8$ and $B=4$.}
     \label{fig:SER_vs_SNR_low}
 \end{figure}

We start with comparing the uncoded symbol error (SER) of the detection methods for different values of the pre-processing SNR=$\frac{P}{\sigma^2_n}$ where $P$ is the expected transmitting power of the users. 
We consider three types of channels according to the correlation matrices introduced in \eqref{eq: BS-U-channel} as 
\begin{enumerate}
    \item Uncorrelated channel with $\mathbf{H}\sim \mathcal{CN}(0,\mathbf{I})$.
    \item Lowly correlated channel with $\theta_i=7\pi/8$, yielding $\bm{R}_i$ very close to the identity matrix.
    \item Highly correlated channel with $\theta_i=3\pi/4$.
\end{enumerate}

In Fig~\ref{fig: sim SER SNR}  we compare the performance of the aforementioned methods for the three types of channels.
First, in 
 an uncorrelated Rayleigh fading channel,
 our VMP based method works very close to the lower bound and clearly outperforms the benchmarks.
For the case of the lowly correlated channel, our method still performs to the matched filter bound. However, for the highly correlated channel, the performance gain over linear methods is small. One way to boost the performance is to add a central ZF initialization to our VMP receiver, which significantly increases its complexity. As expected, this ZF initialization is showing better performance than the local MRC initialization one. 
The results show the superior performance of the proposed distributed VMP method where it performs very close to the ideal bound. Having a SIC mechanism in combination with VMP processing provides the best performance in high SNR regimes where the effect of error propagation becomes negligible. Another observation from Fig.~\ref{fig: sim SER SNR} is the large performance degradation caused by correlated and non-stationary channels compared to an uncorrelated one.
This is due to the channel capacity reduction that has been shown in \cite{gesbert2002outdoor} for the correlated channels and in \cite{li2015capacity} for the non-stationary channels.
The reason for the poor performance of the EP based method is because of the relatively high ratio of the users and antennas $\frac{M}{K}\lessapprox 10$. Furthermore, the complex correlated channel model of \eqref{eq: user k channel} impairs the receiver and makes it perform only slightly better than the centralized MRC.  The EP based method works significantly better in the unrealistic i.i.d. channel model with smaller number of users\cite{wang2019expectation}, as shown in Fig.~\ref{fig:SER_vs_SNR_low}. Here, we compare the methods in an i.i.d channel and a lower system load $\frac{M}{K}= 12.5$. As it can be seen, the EP based method performs much closer to the central ZF curve. For the rest of the simulations we use the lowly correlated channel model.

 \begin{figure}
     \centering
     \includegraphics[width=0.7\linewidth,trim={1.4cm 0 1.6cm 0.6cm },clip]{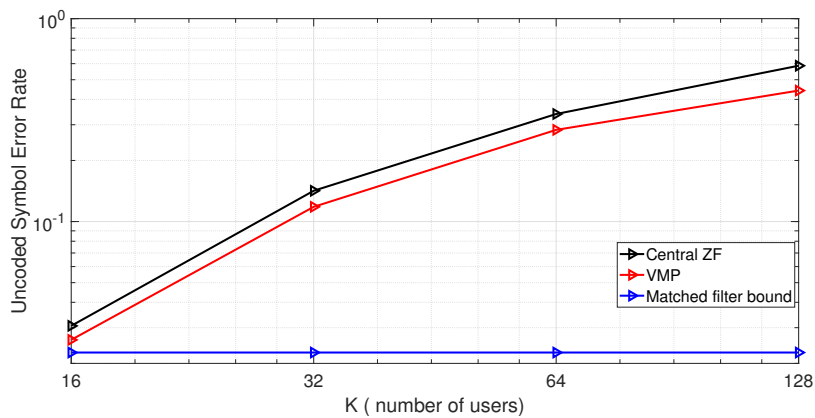}
     \caption{\small Effect of number of the users in performance of the detection methods. The VMP based method degrades slower than the ZF receiver with increasing $K$. (SNR= $10$ dB)}
     \label{fig:SER_vs_K}
 \end{figure}
 
 In Fig.~\ref{fig:SER_vs_K}, we compare the performance of the receivers  with respect to the number of users $K$. The total array size is kept fixed at $M=256$. We can observe that as we add more user in the system and go towards crowded scenarios, the VMP method, which is using the MRC initialization, degrades at a slower rate than the ZF receiver. The reason is the fact that the linear receivers fail to operate properly when the number of users becomes comparable with the number of the antennas at the BS ($\frac{M}{K}<10$). With a larger number of users present, the probability that a user's channel is approximately orthogonal to that of all other users decreases, with large degradation of the ``favorable propagation" conditions usually present in massive MIMO channels. Although the performance also degrades considerably for the VMP receiver, the degradation is less than for the ZF.
 
 \begin{figure}
     \centering
     \includegraphics[width=0.7\linewidth,trim={1.4cm 0 1.6cm 0.6cm },clip]{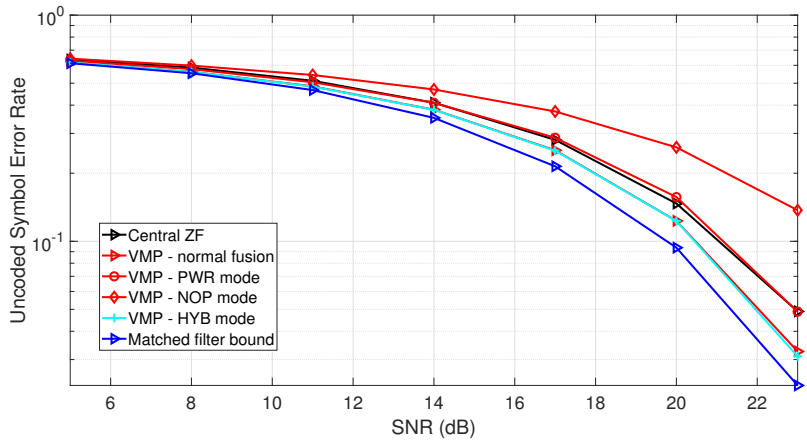}
     \caption{\small The VR based methods and their performance compared to the benchmarks and also the VMP with full sub-array data fusion. $B_{\text{max}}=3$ and $p_0=0.75$.}
     \label{fig:SER_vs_VRbased}
 \end{figure}
 
 The SER of the different VR based modes which restrict data fusion to $75\%$ of the sub-arrays according to section \ref{sec:Data Fusion at the CPU } is illustrated in Fig.~\ref{fig:SER_vs_VRbased}. The VMP method with the fusion of data from all sub-arrays, i.e. with $\mathbf{V}=\mathbf{1}_{B\times K}$, provides the best performance as expected. The power-based data fusion technique (PWR) provides the same performance as the ZF receiver,  while the noise precision based (NOP) receiver performs poorly. 
 The receiver with hybrid (HYB) is the best among VR-based methods, as it approaches the performance of full data fusion while only requiring approximately $75\%$ of its complexity.\footnote{The term ``approximately'' is used because we still need to process $B$ of $\lambda_b$ values at the CPU and only activate the first $B_{\text{max}}$ of the LPUs.} (The complexity reduction is scaled with $\frac{B_{\text{max}}}{B}$) 
 These results illustrate that exploiting non-stationary properties helps to obtain a receiver with lower computational complexity and almost the same performance of a receiver fusing data from all sub-arrays.
 

  \begin{figure}
     \centering
     \includegraphics[width=0.7\linewidth,trim={1.4cm 0 1.6cm 0.6cm },clip]{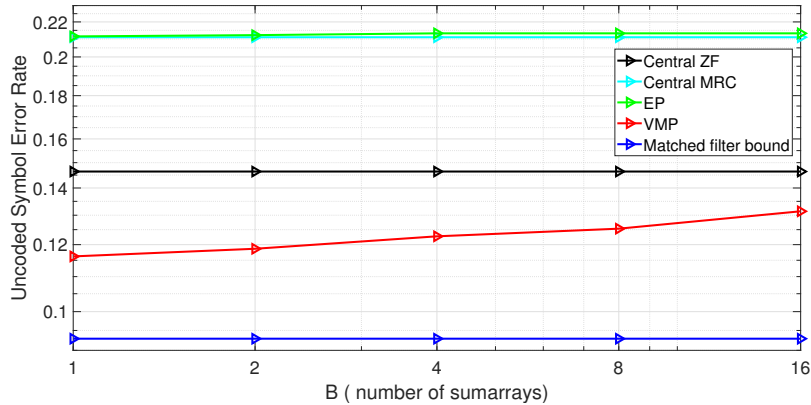}
     \caption{\small Effect of number of the sub-arrays $B$ on the SER behaviour of the distributed schemes. The size of the array $M$ is fixed to $128$.  (SNR= $20$ dB)}
     \label{fig:SA_size}
 \end{figure}
 
 In Fig.~\ref{fig:SA_size} we analyze the performance of the VMP receiver as the processing is distributed among different number sub-arrays. With an array of fixed size $M=128$, the performance at an SNR of 20 dBs is analyzed for different number of subarrays $B$. The central ZF and MRC receivers and the matched filter bound are unaffected by $B$. Predictably, distributing the processing among more local units leads to an increase in the SER, although the performance is still better than that of centralized receivers. Note that in the rightmost point of this figure, with $B=16$, the number of antennas per sub-array is $M_b=M/B=2K=16$ which is an extreme case for a massive MIMO system, but the VMP receiver still operates with acceptable performance. 
 

   \begin{figure}
     \centering
     \includegraphics[width=0.7\linewidth,trim={1.4cm 0 1.6cm 0.6cm },clip]{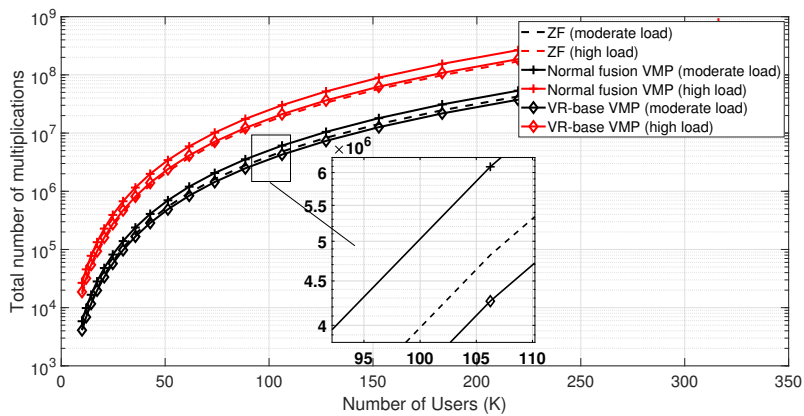}
     \caption{\small Number of complex multiplications for the VMP based methods and the central ZF for different load conditions. For the VR-based VMP we assume to have $\frac{B_{\text{max}}}{B}=0.75$}
     \label{fig:complexity}
 \end{figure}
 
We finalize by showcasing the complexity of the assessed receivers in Fig.~\ref{fig:complexity}. We consider two system load regimes: one  with $M/K=10$ for a moderate load, and $M/K=5$ in a crowded scenario. As it can be observed, the complexity of the VMP method when it fuses all the sub-arrays is higher than the ZF method. As we discussed before, this is a trade-off point where we get a near to optimal performance while spending more computational resources. Moreover, we can see that the complexity of a VR based method with hybrid fusion mode is close to the ZF while it still provides better SER output. 
These results illustrate that the proposed VMP receiver can be tuned to trade-off performance and computational complexity depending on the system requirements and operating conditions.


 \section{Conclusions}
 
 We propose a distributed receiver structure based on VMP that outperforms conventional massive MIMO receivers, especially when operating in spatial non-stationary channels. Numerical results show that the receivers implementing the proposed algorithm perform close to a genie-aided receiver (matched filter bound), even in highly correlated channel conditions.
One of the key components of our method is the internal SIC mechanism which takes advantage of the energy  variations over the VRs of different users. This interference cancellation improves the receiver performance for the users with overlapping VRs. Unlike the conventional linear receivers, our VR-based methods use information about the non-stationarities to limit the  complexity without performance degradation. Our design is versatile in several respects: the distributed manner of all the processing tasks makes it easier for practical deployments of the XL-MIMO systems.  The variety of options for initialization,  data fusion and detection methods gives us several control parameters  allowing for trading off computational complexity for performance  and vice-versa  in the receiver design for different applications.
 Our future research will address lower complexity receivers for higher modulations schemes, extending the detectors with different coding methods and using machine learning techniques to optimize the data fusion and the SIC at the CPU.
 
 \bibliographystyle{IEEEtran}

\end{document}